\documentclass[12pt]{article}
\usepackage{amsmath}
\usepackage{amssymb,amscd}
\usepackage{bm}
\usepackage{wrapfig}
\usepackage{color}
\usepackage{amsmath}
\usepackage{amssymb,amscd}
\usepackage{bm}
\usepackage{amsbsy} 
\usepackage{wrapfig}
\usepackage{theorem}

\newcommand{\beq}{\begin{equation}}
\newcommand{\eeq}{\end{equation}}
\newcommand{\bea}{\begin{eqnarray*}}
\newcommand{\eea}{\end{eqnarray*}}
\newcommand{\beqa}{\begin{eqnarray}}
\newcommand{\eeqa}{\end{eqnarray}}

\newcommand{\calL}{{\cal L}}
\newcommand{\calM}{{\cal M}}
\newcommand{\calN}{{\cal N}}

\newcommand\md{\mathrm{d}}

\begin{document}


\baselineskip 6mm 

\begin{titlepage}
\begin{flushright}
\end{flushright}


\begin{center}
{\Large \bf
On the relation of Lie algebroids to 
\\ \vspace{2mm} constrained systems and their BV/BFV formulation
}
\vskip 1cm
Noriaki Ikeda${}^{a}$
\footnote{E-mail:\ 
nikedaATse.ritsumei.ac.jp}
and 
Thomas Strobl${}^{b}$
\footnote{E-mail:\ 
stroblATmath.univ-lyon1.fr
}

\vskip 0.4cm
{
\it
${}^a$
Department of Mathematical Sciences,
Ritsumeikan University \\
Kusatsu, Shiga 525-8577, Japan \\

\vskip 0.4cm
${}^b$
Institut Camille Jordan,
Universit\'e Claude Bernard Lyon 1 \\
43 boulevard du 11 novembre 1918, 69622 Villeurbanne cedex,
France

}
\vskip 0.4cm

{March 2, 2018}
\vskip 1.6cm

\emph{Dedicated to the 50th birthday of Anton Alekseev}

\vskip 1.6cm

\begin{abstract}
We observe that a system of irreducible, fiber-linear, first class constraints on $T^*M$  is equivalent to the definition of a foliation Lie algebroid over $M$. The BFV formulation of the constrained system is given by the Hamiltonian lift of the Vaintrob description $(E[1],Q)$ of the Lie algebroid to its cotangent bundle $T^*E[1]$. Affine deformations of the constraints are parametrized by the first Lie algebroid cohomology $H^1_Q$ and lead to irreducible constraints also for much more general Lie algebroids such as Dirac structures; the modified BFV function follows by the addition of a representative of the deformation charge.  

Adding a Hamiltonian to the system corresponds to a metric $g$ on $M$. Evolution invariance of the constraint surface introduces a connection $\nabla$ on $E$ and one reobtains the  compatibility of $g$ with $(E,\rho,\nabla)$ found previously in the literature. 
The covariantization of the Hamiltonian to a function on $T^*E[1]$ serves as a BFV-Hamiltonian, iff, in addition, this connection is compatible with the Lie algebroid structure, turning $(E,\rho,[ \cdot , \cdot ],\nabla)$ into a Cartan-Lie algebroid. The BV formulation of the system is obtained from BFV by a (time-dependent) AKSZ procedure.

\end{abstract}
keywords: Constrained systems, Hamiltonian and Lagrangian systems, BFV and BV formalism, AKSZ formalism, Lie algebroids, 
Cartan connections, higher structures.
\end{center}
\end{titlepage}

\newpage

\noindent {\bf 1.} \emph{Constrained systems} appear in the Hamiltonian description of gauge theories \cite{Dirac,Henneaux-Teitelboim}. In the finite dimensional setting, they consist of an $n$-dimensional symplectic manifold $(N,\omega)$, a Hamiltonian function $H \in C^\infty(N)$, and $r$ constraint functions $\Phi_a$, satisfying the following compatibility condition: There need to be functions $\gamma_a^b \in C^\infty(N)$ such that 
\beq \label{Hflow}
\{ H,\Phi_a  \} = \gamma^b_{a} \, \Phi_b \, ,
\eeq
where  $\{ \cdot , \cdot \}$ denotes the Poisson bracket induced by the symplectic form $\omega$. This requirement ensures that 
 $H$ is preserved by the Hamiltonian flow of the constraints, at least on-shell, i.e.~on the constraint surface $C := \{ \Phi_a=0\} \subset N$, defined as the common zero locus of the constraint functions. Equivalently, it ensures that the Hamiltonian flow of $H$, which generates the time evolution of the system, remains on the surface $C$ always. Next, the Poisson brackets among the constraints play an important role. The most interesting situation occurs when they are of the first class, i.e.~when they satisfy
\beq \label{first}
\{ \Phi_a , \Phi_b \} = C^c_{ab} \,\Phi_c
\eeq 
for some functions $C^c_{ab}$ on $N$. So, a first class constrained system consists of the data $(N,\omega, H, (\Phi_a)_{a=1}^r)$ such that the  compatibility conditions \eqref{Hflow}, \eqref{first} hold true. We call it \emph{topological}, if it satisfies $H \vert_C=0$, i.e.~if the Hamiltonian vanishes on the constraint surface.

There is an important question in this context, namely if the set of constraints $(\Phi_a)_{a=1}^r$ is reducible (redundant) or not.  Mathematically, irreducibility---together with a regularity condition to be satisfied by the constraint functions, which we will assume in any case---can be  defined as the property that $\varphi_C^* \left( \md \Phi_1 \wedge \ldots \wedge \md \Phi_r \right)$ is everywhere non-zero, where $\varphi_C \colon C \to N$ is the canonical embedding map of the constraint surface into the original phase space. Irreducibility of the constraints is easily seen to imply that the restriction of the set of functions  $(\gamma^b_{a})_{a,b=1}^r$ and $(C^c_{ab})_{a,b,c=1}^r$ to the constraint surface $C$ is uniquely determined by equations \eqref{Hflow} and \eqref{first}. Also it is an elementary exercise to show that it implies the following equivalence 
\beq  \label{anti}
\mu^a \,\Phi_a = 0 \qquad \Leftrightarrow \qquad \exists \quad \sigma^{ab}=-\sigma^{ba} \quad \mathrm{such} \;  \mathrm{that} \quad \mu^a =  \sigma^{ab}\,\Phi_b \, .
\eeq
 We call two sets of irreducible constraints $(\Phi_a)_{a=1}^r$ and  $(\tilde \Phi_a)_{a=1}^r$ equivalent, moreover, if there exist  functions $M^a_b$ on $N$ such that 
 \beq \label{equiv}
 \tilde \Phi_a = M^a_b \,\Phi_b
 \eeq
 holds true and the matrix $(M^a_b)_{a,b = 1}^r$ is invertible when restricted to  $C$. For an irreducible first class constrained system, we may more generally require the existence of the functions $(\Phi_a)_{a=1}^r$ locally on $N$ only---with a gluing performed by means of \eqref{equiv}. Such a system is topological then iff, in addition, (locally) $H$ can be written as a linear combination of the constraints, $H=\kappa^a \Phi_a$ for some set  $(\kappa^a)_{a=1}^r$ of (locally defined) functions.
 
 \vspace{5mm}
 \newpage
 \noindent {\bf 2.} \emph{Lie algebroids} are an important notion of contemporary geometry. They consist of a vector bundle $E \to M$ together with a bundle morphism $\rho \colon E \to TM$ as well as a Lie algebra $(\Gamma(E), [ \cdot , \cdot ])$ satisfying the Leibniz rule $[s, f \tilde s]=f[s,\tilde s] + \rho(s)f \, \tilde s$ for all $s,\tilde s \in \Gamma(E)$ and $f \in C^\infty(M)$. There is an elegant equivalent way of describing Lie algebroids by means of $\mathbb{Z}$-graded geometry \cite{Vaintrob}: Every graded manifold $\calM$ with local coordinates $(x^i)_{i=1}^n$ and $(\xi^a)_{a=1}^r$ of degree zero and one, respectively, are of the form  $\calM= E[1]$ for some rank $r$ vector bundle $E$, where the shift by one in the brackets indicates that the fiber-linear coordinates on $E$, i.e.~sections in $E^*$, are considered as degree one odd functions.  The most general degree plus one vector field on $\calM$ has the form:
 \beq \label{Q}
 Q = \rho_a^i (x) \xi^a \frac{\partial}{\partial x^i} - \frac{1}{2} C^c_{ab}(x) \xi^a \xi^b  \frac{\partial}{\partial \xi^c} \, .
 \eeq
 Let $e_a$ be a local basis in $E$ dual to the basis corresponding to the coordinates $\xi^a$, then the data entering $Q$ can be used to define an anchor map $\rho$ and a bracket by means of $\rho(e_a) :=  \rho_a^i \partial_i$  and  $[e_a,e_b]:= C^c_{ab} e_c$. Using the (partially non-trivial) transformation properties of the coefficient functions $\rho_a^i$ and $C^c_{ab}$ as induced from (degree-preserving) coordinate changes on $E[1]$, we can extend $\rho$ and the bracket to all sections of $E$ and deduce the Leibniz rule for the bracket. Finally, one verifies that these data satisfy the remaining axioms of a Lie algebroid, iff
 \beq [Q,Q] \equiv 2 Q^2 = 0 \, .\label{Q2}
 \eeq
Such a pair $(\calM,Q)$ is sometimes called an NQ-manifold  \cite{Schwarz}. Identifying functions on $E[1]$ with sections of $\Lambda^\bullet E^*$, $Q$ corresponds to a differential ${}^E \md$ on these  $E$-differential forms, which is a mutual generalization of the de Rham  differential (take $E=TM$) and the Chevalley-Eilenberg differential (a Lie algebroid over a point is a Lie algebra). 

It is a consequence of the Lie algebroid axioms that $\rho$ is a Lie algebra morphism: $[\rho_a,\rho_b]=C^c_{ab} \rho_c$. Regular Lie algebroids are those where the span of these vector fields has a constant rank and they thus correspond to regular foliations of $M$. In particular, a \emph{foliation Lie algebroid} is one where $\rho$ is injective and thus the foliation regular and for every foliation of $M$ there is, vice versa, a unique such a Lie algebroid up to isomorphism.
In general, the base $M$ of a Lie algebroid $E$ carries a singular foliation only. A fairly general class of such foliations can be obtained by the (symplectic) leaves of a Poisson manifold, or, more generally, the (presymplectic) leaves of a Dirac manifold. Dirac manifolds can be defined as the base of a rank $n$ Lie algebroid called a Dirac structure. It is obtained as a maximally isotropic, involutive subbundle of an exact Courant algebroid, cf., e.g., \cite{Severa-Weinstein}. If one drops the maximality condition, one obtains also lower rank Lie algebroids, with a rank smaller than the dimension of $M$. We call them \emph{small Dirac structures}.

\newpage

 
 
 \noindent {\bf 3.} 
 Let  $N=T^*M$, equipped with its canonical symplectic form $\omega= \md x^i \wedge \md p_i$; locally, $(N,\omega)$ is always of this form, but we require this globally now. Then there is a natural grading of functions with respect to fiber-linearity or, equivalently, by the monomial degree in the momenta $p_i$.  We will now show an \emph{equivalence} between 
 topological, 
 irreducible, first class constrained systems with fiber-linear constraints---up to local equivalence \eqref{equiv}, 
 \beq \label{TstarM} 
 \{(T^*M, \omega_{can}, (\Phi_a)_{a=1}^r - \mathrm{linear})\}/\sim
 \eeq and
 rank $r$  foliation Lie algebroids $E\to M$. 

Denote the set of degree $k$ functions by  $C^\infty_k(T^*M)$.  $\Phi_a \in C^\infty_1(T^*M)$ implies
\beq 
\label{lin}
\Phi_a = \rho_a^i(x) \, p_i \, .
\eeq 
Evidently, the Poisson bracket reduce the degree by one: $\{ p_i , x^j \} = \delta_i^j$. Thus the equality \eqref{first} implies $C^c_{ab} \in C^\infty_0(T^*M) \cong C^\infty(M)$, which is uniquely determined due to the irreducibility condition and the fact that the zero-section of $T^*M$ is contained in $C$. This gives the ingredients for the identification with the data of a Lie algebroid in local coordinates. It remains to relate $Q^2=0$ to the properties of the constrained system.
Equation \eqref{first} takes the form $[\rho_a,\rho_b]^i  = C^c_{ab} (x) \rho_c^i$, which is equivalent to $Q^2(x^i)=0$. 
 
 Next we apply \eqref{first} twice to $\{\{\Phi_a,\Phi_b\},\Phi_c\} + \mathrm{cycl}_{abc}=0$, which results in
 \begin{eqnarray}
\left(C_{ab}^e \,C_{ce}^d + \frac{\partial C_{ab}^d}{\partial x^j} \,\rho_c^j 
 + \mbox{cycl}_{abc} \right) \Phi_d =0\, .
\end{eqnarray}
 Now it is time to use the consequence \eqref{anti} of the irreducibility condition on the constraints: from the above identity we may deduce (squared brackets imply antisymmetrization in the intermediary indices)
 \beq
 C_{[ab}^e \,C_{c]e}^d + \rho_{[a}^j \,C_{bc],j}^d 
   = \sigma^{de}_{abc}\, \Phi_e
 \eeq
 for some functions $\sigma_{abc}^{de}$, antisymmetric in the lower as well as in the upper indices. Since the left hand side is of degree zero and the right hand side at least of degree one, necessarily \beq \label{consist} C_{[ab}^e \,C_{c]e}^d + \rho_{[a}^j\, C_{bc],j}^d= 0 \, .
 \eeq
It is now straightforward to verify that this last equation yields precisely the missing part $Q^2(\xi^a)=0$ of the Lie algebroid axioms.

It remains to remark that the equivalence \eqref{equiv} takes care of the equivalence of the two sides to not depend on the choice of a chosen frame. Moreover, if the constraints $(\Phi_a)_{a=1}^r$ are defined globally, the bundle $E$ is trivial. Gluing locally defined constraints as described at the end of {\bf 1.}, we also  obtain non-trivial vector bundle structures $E$. 

Evaluating the  irreducibility of the constraints \eqref{lin} on the zero-section of $T^*M$, one finds that the multi-vector field $\rho_1 \wedge \ldots \wedge \rho_r$ must be non-zero for every $x\in M$. This implies that $E$ is a foliation Lie algebroid. 

 \newpage 
 \noindent {\bf 4.} Given a Lie algebroid, described by means of the odd vector field $Q$ in equation \eqref{Q}, we can always consider an \emph{associated constrained system} \eqref{TstarM}, \eqref{lin}. It will satisfy the first class property \eqref{first}, with the structure functions of the Lie algebroid read off from $Q$. 
 
The constraints now will, however, no more be necessarily irreducible. Assume, for example, that there exist $k$ sections $(s_I)_{I=1}^k$ which are in the kernel of the anchor map, $\rho(s_I)=0$ for all $I=1, \ldots , k$. This is evidently tantamount to the following $k$ reducibility conditions among the constraints:
 \beq s_I^a \, \Phi_a = 0 \qquad \forall I=1, \ldots , k. \label{redun}
 \eeq

If one drops the irreducibility condition of the constrained system  \eqref{TstarM}, in general we can no more reverse the argument for getting a Lie algebroid from these data. It already starts with the fact that due to equations of the form \eqref{redun} one can no more read off the structure functions for a bracket on $E$ by means of equation \eqref{first}, these coefficients are not unique now. It is not even clear if they \emph{can} be chosen to satisfy \eqref{consist}, which is necessary for yielding a Lie algebroid structure on $E$. In fact, this question is equivalent to the one if there exists a Lie algebroid structure generating a singular foliation ${\cal F}$ on its base $M$ (here ${\cal F}$ is the projection from $T^*M$ to $M$ of the foliation generated by first class constraints). In general, globally this is not always the case in the smooth setting and locally the problem is still open, cf, e.g., \cite{CST} for a discussion.

 On the other hand, under a condition like \eqref{redun}, we can introduce a rank $k$ bundle $F$ over $M$ and a bundle map $t \colon F \to E$. If the equations \eqref{redun} parametrize all redundancies between the constraints $(\Phi_a)_{a=1}^r$, then $\Gamma(F) \to \Gamma(E) \to \Gamma(TM)$ becomes an exact sequence. One then may ask oneself, if there is possibly a 3-bracket on this complex, so that the violation of the Jacobi identity for the 2-bracket on $E$ can be written as the image of this 3-bracket taking values in $F$, i.e.\ up to something exact in the complex, or, in other words, up to homotopy. There can be also dependencies between the dependencies: assume that there are functions $(g^I_\alpha)_{\alpha = 1}^m$ such that $ g^I_\alpha \, s_I^a = 0$ $\forall \, \alpha=1,\ldots,m$ and $a=1, \ldots, r$ and then the 3-bracket might be asked to satisfy its higher Jacobi type identity only up to some homotopy again. Evidently such a procedure might be continued and one is led to the realm of higher Lie algebroids or, in the case that the procedure never stops, to Lie infinity algebroids. Such higher Lie algebroids, say a Lie n-algebroid for some $n>1$, can again be described by means of an NQ-manifold, only that now the degrees of local coordinates on the non-negatively graded manifold will be bounded by $n$ instead of by 1.

We thus find that general topological, linearly constrained first class systems \eqref{TstarM}  lead to higher Lie algebroid structures defined over $M$. This is intimately related to \cite{CST}. In particular, every Lie $n$ algebroid over $M$ gives canonically rise to a topologically constrained system of the form \eqref{TstarM}, together with all its redundancy relations needed for the construction of the corresponding BFV formulation with its ghosts for ghosts etc.

 \vspace{5mm}
 \newpage
 \noindent {\bf 5.} A straightforward \emph{generalization} of the system \eqref{TstarM} occurs when one permits the constraints to be polynomials---instead of just monomials---of degree one: $\Phi_a \in C^\infty_{\leq 1}(T^*M)$. In this case we can write 
 \beq  \label{affine}
 \Phi_a= \rho_a^i(x) p_i + \alpha_a(x) \, .
 \eeq 
 Assume that the constraint algebra  \eqref{first} is satisfied for them for structure functions $C^a_{bc}(x)$ which, together with the coefficients  $\rho_a^i(x)$, come from a Lie algebroid on a bundle $E$ over $M$. Then $\alpha_a$ can be considered as the components of an $E$-1-form, $\alpha= \alpha_a(x) e^a$, where $e^a$ is a basis dual to the chosen one on $E$ used in the local coordinate system underlying \eqref{Q}.  The condition \eqref{first} now becomes equivalent to the following natural condition on $\alpha$,  \beq \label{Ed} {}^E\md \alpha = 0 \, . \eeq
 On the other hand, $\alpha$ is determined by \eqref{affine} only up to additions of the form $\alpha_a \mapsto \alpha_a +  \rho_a^i(x) \partial_i f(x)$, since redefining $p_i$ by the gradient of a function on $M$ does not modify the symplectic form. Since such additions to $\alpha$ are the ${}^E\md$-exact ones, we see that affine deformations of first class constraints \eqref{lin} are parametrized by  Q-cohomology of the Lie algebroid at degree one,
 \beq \label{coho}
 [\alpha] \in H_Q^1(E[1]) \, .
 \eeq 
The interest in this construction lies also in the fact that the affine constraints permit to host much more general Lie algebroids. For example, for every Dirac structure, there exist cohomology classes $[\alpha]$ such that the first class constraints \eqref{affine} become irreducible. We leave the proof of this statement to the reader. On the other hand, the (constant) structure functions of Lie algebra bundles or, more generally, the Lie algebra in the kernel of the anchor map, will never be reflected in the Poisson brackets of the associated constraints. So there the relation of Lie algebroids to constrained systems finds its limitations.

We conclude this section with a somewhat more abstract and in particular coordinate independent reformulation of the constraints. For a fixed Lie algebroid $E$, the constraints \eqref{affine} can be seen as a Lie algebra morphism
\beq \Phi \colon (\Gamma(E),[\cdot,\cdot]) \to (C^\infty_{\leq 1}(T^*M), \{ \cdot , \cdot \})\; , \; \: s \mapsto \rho(s) + \langle \alpha,s\rangle\, . \label{Phi}
\eeq
Here the vector field $\rho(s) \in \Gamma(TM)$ and the function $\langle \alpha,s\rangle \in C^\infty(M)$ are considered as functions of degree one and zero on $T^*M$, respectively. Both Lie algebras in \eqref{Phi} are acted upon by the ring of functions $C^\infty(M)$.  The validity of the corresponding Leibniz rules lifts the above morphism to a morphism of Lie-Rinehart algebras \cite{LieRinehart}. In this language, the constraints are irreducible, iff the zero level set of the image of $\Phi$ is a codimension $r$ coisotropic submanifold in $T^*M$. For linear constraints, $\alpha=0$, the existence of reducibility conditions of the form \eqref{redun} is equivalent to  the injectivity of the map \eqref{Phi}. In the affine case, the situation is more intricate and what counts for irreducibility is injectivity of the map $\Phi$  when restricted to a particular subset of sections in $E$.

 \vspace{5mm}
 
 \noindent {\bf 6.} We obtain a non-topological theory, if we add a \emph{Hamiltonian} $H$ to the system \eqref{TstarM}. The physically most relevant case is where $H\in C^\infty_2(T^*M)$, 
\beq\label{H}
H= \frac{1}{2} g^{ij}(x) p_ip_j\, , 
\eeq
 and the matrix $g^{ij}$ has an inverse corresponding to a metric $g$ on $M$. We now want to analyze 
 what the consistency condition \eqref{Hflow}, standard within the context of constrained systems, translates into 
 for the metric on the base $M$ of the corresponding Lie algebroid $E$---for simplicity, we again assume irreducibility of the constraints here so that $E$ is a foliation Lie algebroid. For degree reasons,   $\gamma^a_b$ needs to be linear in the momenta: $\gamma^a_b= \gamma^{ai}_b(x)\, p_i$. Moreover, it is obvious from equation  \eqref{Hflow} that these coefficients do not transform like a tensor with respect to a change of basis \eqref{equiv}; introducing $\omega^a_{bi} = g_{ij} \gamma^{aj}_b$, $\omega^a_b\equiv \omega^a_{bi} (x) \md x^i$ transforms instead precisely like the 1-forms of a connection $\nabla$ on $E$. 
 
Thus the compatibility condition \eqref{Hflow}  will translate into a condition between the metric $g$ on $M$, the anchor $\rho$ of the bundle $E$, and the connection $\nabla$. In fact, $\nabla$ and $\rho$ can be combined to define an 
$E$-connection  on $TM$, i.e.~a connection where one derives  vector fields $v$ on $M$ along sections $s$ of $E$: ${}^E\nabla_{\! s} v := \calL_{\rho(s)} v + \rho(\nabla_v s)$. The dynamical consistency of the constrained system now turns into the geometrical compatibility equation
\beq {}^E\nabla g = 0\, ,\label{ginv'}
\eeq
which has to hold true for \emph{some} connection $\nabla$. 
The condition \eqref{ginv'} was already obtained in the context of gauging sigma models in $d+1$ spacetime dimensions within the Lagrangian formalism  \cite{withoutsymmetries}. The consideration above is the Hamiltonian counterpart of that analysis for $d=0$ and reinforces the observation made there from yet this second point of view: Rewriting \eqref{ginv'} into the equivalent ${\cal L}_{\rho_a} g = 
\omega_a^b  \vee \iota_{\rho_b} g$, one sees that the vector fields $\rho_a$ do not need to be Killing vectors of the metric for the existence of a gauge theory---here: for the consistency of the  first class constrained system. The Hamiltonian equation \eqref{Hflow}  explains well the heuristic meaning: When going along the leaves of the foliation, only those components of the (inverse) metric which go along the leaves may change. We refer to  \cite{KS16,KS18} for a purely geometrical analysis of this and similar compatibility equations; there it is shown, e.g., that  for an arbitrary Lie algebroid the condition \eqref{ginv'} implies that the (possibly singular) foliation on $M$ induced by $(E,\rho)$ is  Riemannian with respect to $g$.

It was one of Dirac's brilliant ideas \cite{Dirac} that, in a constrained Hamiltonian system, it is advisable to replace the original Hamiltonian $H$ by its extension
\beq H^{ext}(x^i,p_i,\lambda^a) := H(x,p) + \lambda^a \Phi_a(x,p) \, ,\label{ext}
\eeq
where $\lambda^a$ are arbitrary Lagrange multiplier coordinates. 
Choosing them to be linear in $p$, we see yet from another perspective that $g^{-1}$ can be changed arbitrarily along the directions generated by $\rho$; only its transversal components are  physical or of relevance.


 \vspace{5mm}
 
 \noindent {\bf 7.} Let us now consider a combination of the last two items, generalizing simultaneously the Hamiltonian to also be a \emph{polynomial}: $ \Phi_a \in C^\infty_{\leq 1}(T^*M)$ and $H \in C^\infty_{\leq 2}(T^*M)$.
Thus the constraints $\Phi_a$ are of the form \eqref{affine}  and $H$ can be parametrized according to 
\beq \label{H2}
H= \frac{1}{2} g^{ij}(x) p_ip_j+ \beta^i(x)p_i + V(x) \, .
\eeq 

We can get rid of the term linear in the momenta, $\beta^i \mapsto 0$, at the expense of redefining the potential $V$ and the $E$-1-forms $\alpha$ and simultaneously twisting the symplectic form $\omega_{can}$ by a magnetic field $B=\md A \in \Omega^2(M) \subset \Omega^2(T^*M)$, were $A=A_i(x) \md x^i$ and $A_i = g_{ij} \beta^j$:
\beq \omega = \omega_{can} + B \, . \label{B}
\eeq
 Permitting $B$ of non-trivial cohomology, corresponds to a Wess-Zumino term in $d=0+1$.

Due to $\{ p_i,p_j\} = B_{ij}$, the first class property of the affine constraints now requires 
\beq {}^E \md \alpha = \rho^*(B) \, . \label{alpha'}
\eeq 
Here $\rho^*$  
is the map induced by the dual of the anchor, mapping ordinary differential forms to $E$-differential forms. 
In particular, $\rho^*(B)= \frac{1}{2} B_{ij} \rho_a^i \rho_b^j \xi^a \xi^b \in \Gamma(\Lambda^2E^*)\equiv{}^E\Omega^2(M)$. 
It is a nice exercise to show that $\rho^*$ is a chain map, thus descending to cohomology, $\rho^* \colon H^p_{deRham} \to H^p_Q(E[1])$. Equation  \eqref{alpha'} then implies that the choice of the deRham cohomology class of $B$ is restricted  to lie in the kernel of $\rho^*$. Otherwise the  associated constrained system is obstructed to be of the first class. For every permitted choice of $B$, however, like for the original $B=\md A$, equation \eqref{alpha'} shows that  the affine deformations of the constraints form again a torsor over $H^1_Q(E [1])$.


Let us now consider the dynamics. The consistency condition \eqref{Hflow} now not only introduces a connection on $E$, but in addition also a section $\tau \in \Gamma(\mathrm{End}(E))$: 
\beq \gamma^{a}_b =
\omega^a_{bi} g^{ij }p_j + \tau^a_b \, .
\eeq
Equation \eqref{Hflow} then gives three conditions by considering it to second, first, and zeroth order in the momenta. To second order, we find the unaltered condition  \eqref{ginv'} on the metric $g$. To first order we get a constraint on the system of constraints, in addition to \eqref{alpha'}: It relates the exterior covariant derivative of $\alpha$ induced by $\nabla$, $\mathrm{D}\alpha \in \Gamma(E^* \otimes T^*M)$, to the anchor map $\rho$, now viewed upon as a section of $E^* \otimes TM$:
\beq \mathrm{D}\alpha + (\tau^t \otimes g_\flat) \rho = 0 \, ,
\eeq 
where $\tau^t \colon E^* \to E ^*$, the transposed of $\tau$, and $g_\flat \colon TM \to T^*M, v \mapsto \iota_vg$, as maps on the corresponding sections. And 
to zeroth order one finds that the potential $V$ has to satisfy
\beq {}^E\md V = \tau(\alpha) \, .
\eeq
In other words, $\tau$ is seen to govern the violation of covariant constancy of $\alpha$ and, simultaneously, 
the deviation of $V$ to be a Casimir function on $M$. 

 \vspace{5mm}
 \newpage
 \noindent {\bf 8.} We now turn to the \emph{ BFV formulation} \cite{BFV1,BFV2} of the constrained system. Consider $\calN= T^*(E[1])$ for some Lie algebroid $E$. As every cotangent bundle, it carries a symplectic form $\omega_\calN$. Also  there is a canonical lift of vector fields on the base to Hamiltonian functions: one just reinterprets the sections of the tangent bundle as (fiber-linear) functions on the cotangent bundle ($v^i(x)\partial_i \cong v_{Ham} \equiv v^i(x)p_i$). For the vector field  \eqref{Q} on $E[1]$  this yields \beq \label{Qham}
 Q_{Ham} = \rho_a^i (x) \xi^a p_i - \frac{1}{2} C^c_{ab}(x) \xi^a \xi^b  \pi_c \, ,
 \eeq 
where $p_i$ and $\pi_a$ are the momenta conjugate to $x^i$ and $\xi^a$, respectively. The momenta carry the same ghost degree as the derivatives they stand for, i.e.~0 and -1, respectively.
Due to \eqref{Q2} and the fact that there are no degree 2 constants, \eqref{Qham} satisfies $ \{ Q_{Ham}, Q_{Ham}\}_\calN = 0$.  
The BFV formulation of an irreducible topological constrained system \eqref{TstarM} is then simply 
\beq \label{BFV} (\calN_{BFV}= T^*E[1] \, , \; S_{BFV} = Q_{Ham} \, , \; (\cdot , \cdot)_{BFV} = \{ \cdot , \cdot \}_\calN) \, .
\eeq
The classical master equation of the constrained system is satisfied by construction:
\beq \label{master} ( S_{BFV}, S_{BFV} )_{BFV} = 0 \, .\eeq
This equation also holds true for the case that one considers the constrained system \eqref{TstarM} for an arbitrary Lie algebroid $E$, cf.~{\bf 4.}~above. However, for $E$ non-foliation, \eqref{Qham} is not a valid BFV function. Whenever the reduced phase space $N^{red}$, i.e.\ the quotient of $C$ by the foliation generated by the constraints,  is smooth, one requires an isomorphism 
\beq H_{BFV}^0 \cong C^\infty(N^{red}) \label{iso}\eeq 
Here $H_{BFV}^0$ denotes the $BFV$-cohomology at degree 0. $C^\infty(N^{red})$ contains the ''physical observables''.
E.g.,  \eqref{Hflow} ensures that $H$ descends to  $N^{red}$, it is interpreted as the ''energy'' of the system.
In physical applications, $N^{red}$ is almost never smooth and even if it were, it would not permit a useful direct description. Thus, one prefers to work with the redundant description of the original constrained system or the cohomological approach of the enlarged B(F)V phase space. This latter point of view then proves particularly powerful for quantization. Even if in general $N^{red}$ is not smooth, the validity of \eqref{iso} for every smooth $N^{red}$  is the lithmus test for the physical acceptability of the BFV formulation. 

Lie algebroids with redundancies of their constraints such as in \eqref{redun} fail this test when using \eqref{Qham}. However, under some technical assumptions, they can be reinterpreted also as higher Lie algebroids with vanishing higher brackets, but over a non-trivial complex. For higher Lie algebroids the description in terms of an NQ-manifold $(\calM,Q)$ introduces further graded variables, where $\calM$ is now strictly bigger than $E[1]$. Then indeed, $S_{BFV}$ can again be identified with the Hamiltonian  $Q_{Ham}$  on $\calN = T^*\calM$. This shall be made explicit elsewhere.
Interestingly, also on the purely mathematical side, higher Lie algebroids are more
useful than ordinary ones for 
 finding invariants associated to a singular foliation, even if this foliation happens to be generated by a normal Lie algebroid, cf.~\cite{CST}.

 \vspace{5mm}
 \newpage
 \noindent {\bf 9.} Let us now look at the  \emph{non-topological generalization} of the previous item. For this we need to find the BFV-extension of the Hamiltonian function $H$ induced by a metric. We first observe that  \eqref{H} does not define a function on our BFV-phase space \eqref{BFV} yet. The reason is easy to find: the vector fields $\partial_i$ on $M$ need a connection $\nabla$ to be lifted to $E\to M$. This leads to the covariant derivative $\nabla_i \equiv \partial_i - \omega_{ai}^b \xi^a \partial_a$ instead, where $\xi^a$ corresponds to a basis of (local) sections on $E^*$, viewed as fiber coordinates on $E$. Viewing, for each fixed index $i$, the corresponding vector field on $E$ as a function on $T^*E$---or its shifted version $T^*(E[1])$---we obtain the covariantized momenta
 \begin{equation} p^{\!\nabla}_{\:i} := p_i -\omega_{ai}^b \xi^a \pi_b \, .
 \end{equation}
 Equivalently, one is led to such an expression by regarding the transformation of the momenta in $\omega_{BFV}=\md x^i \wedge \md p_i + \md \xi^a \wedge \md \pi_a$ with respect to the point transformations $x^i \mapsto x^i$, $\xi^a \mapsto M^a_b(x) \xi^a$.  Thus the covariantization of the Hamiltonian \eqref{H} is simply
\beq\label{HBFV}
H_{cov}= \frac{1}{2} g^{ij}(x) p^{\!\nabla}_{\:i}p^{\!\nabla}_{\:j} \qquad \in C^\infty(T^*(E[1]))\, .
\eeq
Note that this is an expression of total degree zero, but a polynomial of ghost degree  two. To serve as the BFV-extension of $H$, this expression needs to be BFV-invariant.  A somewhat lengthy but straightforward calculation, with which we do not want to bore the reader, shows, however, that, without any further conditions than those found already before, one has 
\beq (S_{BFV},H_{cov}) =  -g^{ij}p^{\!\nabla}_{\:i} S_{jab}^c \xi^a\xi^b\pi_c \, .
\eeq 
Here $S_{jab}^c$ are the components of a tensor $S \in \Gamma(T^*M \otimes E \otimes \Lambda^2E^*)$ that was found first in \cite{StroblMayer}. It was shown in \cite{KS16} that $S$ vanishes, \emph{iff} the connection $\nabla$ is compatible with the Lie algebroid structure on $E$, i.e.~iff $(E,\rho,[\cdot, \cdot], \nabla)$ is a \emph{Cartan}-Lie algebroid \cite{Blaom}. 

We can now ask for the BFV reformulation of the constrained system in the case that the constraints and the Hamiltonian are only affine, see the formulas  \eqref{affine} and \eqref{H2}. Again we only present the result of this investigation. In the case that all the structural equations found already before are satisfied and that the connection satisfies the Cartan compatibility condition \cite{Blaom},  the non-topological BFV-formulation of the constrained system takes the form:
\begin{eqnarray}\label{BFV2} \calN_{BFV}&=& T^*E[1] \, , \; \nonumber \\
\omega_{BFV}&=&\md x^i \wedge \md p_i + B + \md \xi^a \wedge \md \pi_a
 \; , 
\nonumber \\ \qquad S_{BFV} &=& Q_{Ham} + \alpha \, , \; \nonumber \\ H_{BFV} &=& H_{cov} + V  \, .
\label{BFVform}
\end{eqnarray}
Here $\alpha=\alpha_a \xi^a$ and the 2-form $B$ on $M$ is canonically pulled back to a 2-form on $T^*(E[1])$. 
 \vspace{5mm}
 
 \noindent {\bf 10.} The BFV form of a physical theory is the Hamiltonian counterpart of its \emph{BV form}  \cite{BV1,BV2}, which is related to the Lagrangian formulation of the theory. There exist general formulas, cf., e.g., \cite{Henneaux-Teitelboim}, of how to obtain the BV formulation for a given set of BFV data, in our case the formulas \eqref{BFVform} above. Most elegantly \cite{Grigoriev-Damgaard},  the transition from BFV to BV can be performed as an AKSZ procedure \cite{AKSZ} (see also \cite{Cattaneo-Felder-AKSZ,Roytenberg,IkedaAKSZ} for AKSZ and \cite{BBD1,BBD2,Grigorievplus} and references therein for the present context).
 
 Recall that in the AKSZ formalism, one needs a source $Q$-manifold ${\cal S}$ with an integration measure. In our case this is ${\cal S}=T[1]\mathbb{R}\ni (t,\theta)$, the super-time manifold, together with its deRham differential $\md_\mathbb{R} \equiv \theta \frac{d}{dt}$ and standard integration. To avoid boundary contributions, one may use ${\cal S}=T[1]\mathbb{S}^1$ or impose appropriate boundary conditions. 

The target ${\cal T}$, on the other hand, is a PQ-manifold, i.e.~a $Q$-manifold with compatible, homogeneous symplectic form $\Omega$; this means, in particular, that the odd, degree plus one vector field $Q$ has a Hamiltonian function ${\cal Q}$, $Q=\{ {\cal Q} , \cdot \}_\Omega$. If $\Omega$ has non-zero degree, it automatically has a symplectic potential, otherwise we need to impose it---or deal with Wess-Zumino terms. In our case, we choose $\Omega= \omega_{BFV}$, which, for exact magnetic field $B=\md A$, also has a symplectic potential; for simplicity, we put $B$ to zero from now on. In some physical systems one needs to consider time-dependent Hamiltonians. In our case, this applies to the Hamiltonian ${\cal Q}$, which depends at least on the odd part $\theta$ of super-time:
\beq {\cal Q} = S_{BFV} + \theta \, H_{BFV} \, ,
\eeq
a combination \cite{BBD1,BBD2}, which is nilpotent with respect to the BFV-bracket by construction. 

Now, the BV manifold ${\cal N}_{BV}$ is the mapping space of all (not necessarily degree-preserving) supermaps from the source ${\cal S}$ to the target ${\cal T}$, i.e.
\beq {\cal N}_{BV} := \underline{\mathrm{Hom}}(T[1]\mathbb{R}, \calN_{BFV}) \, .
\eeq
The degree minus one BV symplectic form is obtained from the degree zero BFV form as follows,
\beq \omega_{BV} = \int_{T[1]\mathbb{R}} \md t \wedge \md \theta \: \mathrm{ev}^*(\omega_{BFV}) \, . \eeq 
Here we used the evaluation map $\mathrm{ev} \colon {\cal S} \times {\cal N}_{BV} \to {\cal T}$.
 Denoting the superfields on super-time by corresponding capital letters, so that, e.g., $P_i(t,\theta) \equiv p_i(t) + \theta \,  p^{odd}_i(t)$, the BV-functional now takes the typical AKSZ form:
\beq S_{BV} = \int_{T[1]\mathbb{R}} \md t \wedge \md \theta \left(P_i \md_\mathbb{R} X^i - \Pi_a \md_\mathbb{R} \Xi^a -\mathrm{ev}^*{\cal Q} \right)
\eeq
 Note that this ''AKSZ-type'' theory is far from always topological; in fact, we just showed that we can describe \emph{every} physical system with a Hamiltonian formulation in this way. The target Hamiltonian ${\cal Q}$ is time-dependent in general and even for a conservative system still depends on time's super-partner $\theta$.

 \vspace{5mm}
 
 \noindent {\bf 11.} Much of what we wrote 
 can be generalized straightforwardly to \emph{field theories} defined over a space $\Sigma$ of arbitrary dimension $d$, the discussion up to here corresponding to $d=0$. Assume, for example, that one has a first class constrained system of some topological field theory with the constraints satisfying
 \beq \label{firstfield}
\{ \Phi_a(\sigma) , \Phi_b(\sigma') \} = \delta(\sigma-\sigma')C^c_{ab}(X(\sigma)) \,\Phi_c(\sigma)
 \eeq 
where $\sigma,\sigma' \in \Sigma$ and $X \colon \Sigma \to M$ is part of the phase space variables for some target space manifold $M$. The mere fact we need to know is that $C^c_{ab}(x)$ are structure functions of \emph{some} Lie algebroid $E$ over $M$ in some local choice of basis $e_a$ of sections of $E$ \emph{and} that for every $f\in C^\infty(M)$, one has
\beq \label{anchorfield} \{ \Phi_a(\sigma), f(X(\sigma'))\} = \delta(\sigma-\sigma') \, (\rho_af)(X(\sigma))
\eeq
where $\rho_a \in \Gamma(TM)$ are the vector fields that one obtains when evaluating the anchor map $\rho \colon E\to TM$ on the basis vectors $e_a$. Then, the minimal BFV-functional
\beq \label{SBFVfield} S_{BFV} = \int_\Sigma d \sigma \left(
\Phi_a(x) \xi^a(\sigma) - \frac{1}{2} C_{ab}^c(x) \xi^a \xi^b \pi_c(\sigma) \right), \eeq
satisfies the classical master equation \eqref{master} with respect to the original field-theoretic Poisson algebra
being extended by $\{\xi^a(\sigma), \pi_b(\sigma^{\prime})\} = \delta^a{}_b \delta(\sigma - \sigma^{\prime})$. If the constraints $\Phi_a$ are irreducible and the theory topological, moreover, it provides the physical BFV theory.

An example of these considerations is provided by the constraints $J_s[\varphi]$ on the cotangent bundle of loop space as considered in \cite{Alekseev-Strobl} (see also \cite{Zabzine1,Zabzine2,Ikeda+} for some generalizations). They are 
labeled by sections $s$ in a (possibly small) Dirac structure as well as by test functions $\varphi \in C^\infty(S^1)$. This system satisfies in particular the conditions  \eqref{firstfield} and \eqref{anchorfield}. For example, on finds that 
\beq \{ J_s[\varphi],  J_{s'}[\varphi'] \} =  J_{[s,s']}[\varphi\varphi'] \, ,
\eeq
where $[s,s']$ denotes the Courant-Dorfman bracket \cite{Dorfman1,Courant} twisted 
by $H$ \cite{Severa-Weinstein}, which, when restricted to a (possibly small) Dirac structure becomes the Lie bracket of a Lie algebroid. For Dirac structures projectable to $TM$, the BFV-functional takes the minimal form \eqref{SBFVfield}---with $\Phi_a$ replaced by $J_a$, resulting from choosing a local basis $s_a$ in the (small) Dirac structure. Otherwise there are finitely many dependences and one needs further global ghosts to satisfy the test \eqref{iso} in smooth cases, cf., e.g., \cite{Schaller-Strobl}.

A Lagrangian leading to this system is the Dirac sigma model \cite{DSM}. It can be also non-topological \cite{universal1,universal2}, in which case there is a non-vanishing Hamiltonian. This leads to modifications  similar to what one finds for $d=0$ and shall be discussed elsewhere. 

The transition to the BV-formulation can be performed again as described in the previous item. However, for $d>0$  the target space ${\cal T}={\cal N}_{BFV}$ is infinite-dimensional. But, more importantly, the space-time covariance of the BV functional obtained by this method is not guaranteed. We will illustrate this fact at the example of the twisted Poisson sigma model \cite{Klimcik-Strobl} in \cite{Ikeda-Strobl}.

\section*{Acknowledgments}

T.S.\ wants to thank Anton Alekseev for a long lasting and multiply inspiring friendship. 

\noindent N.I.\ thanks  Anton Alekseev and the university of Geneva for the permission of his staying as a visiting scientist and their hospitality. 

\noindent  We gratefully acknowledge the interest and critical and important feedback of Albin Grataloup and Sylvain Lavau  on earlier versions of this paper.  We also thank Camille Laurent-Gengoux for remarks on the manuscript and Maxim Grigoriev for drawing our attention to the references \cite{Grigoriev-Damgaard,Grigorievplus} and \cite{Barnich}.

\noindent This work was supported by the project MODFLAT of the European Research Council (ERC) and the NCCR SwissMAP of the Swiss National Science Foundation.

\section*{Note added}  A complementary observation to the topic of this paper is that there is a canonical Lie algebroid structure defined over every coisotropic submanifold $C$ \cite{Cattaneo-Felder} or, infinite-dimensional on the Lagrangian level, over the space of solutions to the field equations  \cite{Barnich}. 



%

\end{document}